# Econophysics deserves a revamping

*Paolo Magrassi*

v 1.1 January, 2020

**Abstract:** The paper argues that attracting more economists and adopting a more-precise definition of dynamic complexity might help econophysics acquire more attention in the economics community and bring new lymph to economic research. It may be necessary to concentrate less on the applications than on the basics of economic complexity, beginning with expansion and deepening of the study of small systems with few interacting components, while until thus far complexity has been assumed to be a prerogative of complicated systems only. It is possible that without a thorough analysis at that level, the understanding of systems that are at the same time complex and complicated will continue to elude economics and econophysics research altogether. To that purpose, the paper initiates and frames a definition of dynamic complexity grounded on the concept of non-linear dynamical system.

## 1 Introduction

Stanislaw Ulam was quoted as saying that using the term "non-linear science" is like «calling the bulk of zoology the study of non-elephants» (Campbell, 1985). It is indeed rare, and likely impossible, to come across a natural dynamical system that be exactly linear. And yet science and its applications could not possibly have achieved their great successes without making simplifying assumptions, in order to make most problems tractable. Mathematical models are idealizations of nature, involving abstract entities and, quite often, simplifications.

Linearity is one of those. Linear systems do not exist, just like perfect circles, perfect triangles or straight lines. They are idealizations: mathematical models that use linear operators, i.e. mappings that preserve addition and multiplication. Whenever the effects of the natural non-linearity of things can be considered negligible for the contingent purpose, a mathematical model is built that represents the system as if it were linear: a first-order approximation. An audio amplifier is intrinsically non-linear and its components do not obey the superposition principle (Feynman et al., 1964). Still, within certain amplitude and frequency limits, the circuit will behave in a linear fashion and be useful for hi-fi.

Linear models are also useful because, subject to the hypothesis of linearity, many natural systems resemble one another: their behavior can be described with the same equations even if the contexts are very different, such as mechanics, electronics, chemistry, biology, economics, and so on. On the contrary, non-linear systems each have their own mathematical formalization and often not even that: equations are substituted by numerical simulations. It was indeed only when digital computers started allowing to venture into non-linear territory, that "non-linear science", a.k.a. "complexity science", was born.

In the early 1960s, it was discovered that *finite* variations of future states may originate from *infinitesimal* variations in the initial conditions of a deterministic system (Lorenz, 1963) and that consequently making long-term predictions is impossible in principle. In the 1970s, it became apparent that deterministic systems can exhibit behaviors that cannot be explained on the grounds of the laws governing the components (Anderson, 1972), and that chaotic behavior is a possibility even for conservative systems such as those described by Hamiltonians (Li and Yorke, 1975; May, 1976; Feigenbaum, 1978; Coullet and Tresser, 1978; Ford, 1982).

Some of these findings had already attracted economists (Benhabib and Nishimura, 1979; Benhabib and Day, 1980; Day, 1983; Grandmont, 1983, 1985; Begg, 1984; Boldrin and Montrucchio, 1986;

Van Der Ploeg, 1986) when "econophysics" was born, around 1990. The subsequent "physics invasion" in economics research has brought new methods, techniques and tools and some useful practical results but it has so far failed to achieve the key intended objectives, like reducing «some perplexing results of economic theory [...] to a few elegant general principles with the help of some serious mathematics borrowed from the study of disordered materials» (Săvoiu and Simăn, 2013), or answering questions such as «how large-scale patterns of catastrophic nature might evolve from a series of interactions on the smallest and increasingly larger scales, where the rules for the interactions are presumed identifiable and known» (ETH Zürich, 2019).

Furthermore, the econophysics endeavour has been entirely conducted by physicists (with the aid by some mathematicians and information theorists), and it has attracted very few economists. This compounds with the limited scientific success to create a general atmosphere of disillusion and in fact, more often, disregard by economists (Omerod, 2016).

At about the same time, that is since the 1990s, the natural sciences have worked to understand the complexity challenge that stems from dropping linear assumptions when modelling dynamical systems. Contrary to a widespread belief, this process is far from concluded, and some of its results still have not made it to the whole of physics itself. Economy and finance seem characterized by a large degree of such "complexity", so perhaps econophysics will acquire new lymph when it reasons more deeply about the fundamentals of non-linearity, while at the same time attracting more economists to its ranks.

## 2 Linearity

A dynamical system is "a particle or ensemble of particles whose state varies over time and thus obeys differential equations involving time derivatives" (Nature.com, 2019). A *linear* such system is a set of non-interacting parts / particles: the sole interactions are those of the whole system with its environment.

Furthermore, the linear system's response caused by two or more stimuli is presumed to be the sum of the responses which would be caused by every individual stimulus. In the most general definition, such superposition principle (Feynman et al., 1964) also subsumes homogeneity, meaning that if the system's input is multiplied / divided by some quantity the output will increase / decrease by the same measure. Qualitatively, one can say that in a homogeneous system a modification to the components is proportionally reflected in a modification of the whole:

$$F(a\,x + b\,y) = a\,f(x) + b\,f(y)$$

A linear system / problem can be broken into a sum of mutually independent sub-problems. When, to the contrary, the various components / aspects of a problem interact with each other so as to render impossible their separation for solving the problem step by step or in blocks, then the situation is non-linear or "complex".

The systems and the problems that are encountered in nature are non-linear, and indeed the best definition of the term system is possibly that of "a set of parts that, when acting as a whole, produces effects that the individual parts cannot" (Minati and Pessa, 2006). However, to simplify the studies or for application purposes, one often resorts to linearity as a first-order approximation: if the effects of non-linearity can be considered negligible, a mathematical model can be built that represents the system as if it were linear. This approach is fecund in many situations. As an example: an audio amplifier is intrinsically non-linear: resistance, capacitance and inductance are not concentrated in discrete components and, for that matter, resistors, capacitors, inductors or transistors do not obey the superposition principle. Still, within certain amplitude and frequency limits, it will behave in a linear fashion and be useful for hi-fi; hence, its description throughout audio literature will always be that of a linear system, even if in principle it is not.

Linear models are also useful because subject to the hypothesis of linearity many natural systems resemble one another: their behavior can be described with the same equations even if the contexts are very different, such as mechanics, electronics, chemistry, biology, economics, and so on. A linear oscillator is a model described by the same mathematical equation, whether it be a metal spring, an electric circuit or a stand-alone El Niño. (Complex systems, on the contrary, each have their own mathematical formalization and, in many cases, not even that: equations are substituted by computer simulations.)

Gigantic scientific and technological advances have been made using simplifying linearity assumptions, before computers started allowing to venture into non-linear territory. This is how "complexity science" was born.

### 3 Non-linearity

There had been, in fact, several explorations of non-linear territory made by scholars since the 19$^{th}$ century. H. Poincaré was the first to describe how an apparently simple system subject to deterministic laws, such as that composed of three orbiting celestial bodies (e.g., Sun, Earth and Moon), can exhibit a complex (chaotic) behavior (Poincaré, 1890). Other scholars, including A. Lyapunov, A. Bogdanov, V. Volterra, N. Wiener and W. Weaver, made advances and contributed creating complex system thinking in the first half of the 20$^{th}$ century (Magrassi, 2009).

However, the field has acquired new lymph only with the advent of electronic computers, as they allow to simulate whenever mathematics does not do the job because equations are unknown or intractable. The first application of a digital computer to non-linear exploration gave raise to the Fermi-Pasta-Ulam problem / paradox in the early 1950's (Fermi et al., 1955; Dauxois and Ruffo, 2008), an experiment that can be said to have inaugurated non-linear physics, and that «did much to prepare the fertile soil in which the Kolmogorov – Arnold – Moser theorem and non-linear dynamics / chaos eventually grew» (Ford, 1992).

### 3.1 Non-linearity and non-determinism

Fundamental, then, was the work of mathematician and climatologist Edward Lorenz, who made apparent and fully appreciated the problem that Poincaré had touched upon in his three-body system: when observing the evolution of a complex system (i.e., its trajectory in state space), *finite* variations may originate from *infinitesimal* variations in the initial conditions (Lorenz, 1963). In other words, it becomes effectively impossible to distinguish between two beginnings, even if they are infinitely similar, because the future evolution of the system can differ substantially in the two cases. If we were to model the evolution of continental weather, it may make a difference, to the effect of a probability of a tornado in Texas, whether or not «a butterfly flaps its wings in Brazil». Making long-term forecasts is, in principle, impossible.

### 3.2 Non-linearity and deterministic chaos

A striking exemplification of the above is to be found in another small system such as the one composed of a population of predator animals, a population of preys and the food available to the latter. A linear model turns out simplistic and inadequate for the situation: the population of preys is a function of the predators' population but, in turn, the latter will expand and contract based on the availability of preys and that of vegetable food, which influences the preys population. The "preys – predators – food" system is intrinsically non-linear: none of its components, or couples thereof, may be studied in isolation from the others. And indeed, the Lotka-Volterra equations are a classical example of simple non-linear model of an ecological situation, taking feedback into account.

Purposely simplified, the Lotka-Volterra model leads to the formulation of the so-called logistic map (May, 1976):

$$x_{n+1} = r\, x_n\, (1 - x_n)$$

with $x \in (0,1)$ a number representing the ratio between the current population and the maximum possible in the $n^{th}$ year. Iterating the logistic map it becomes plastically evident how the dynamical value $x_{n+1}$ be always dependent on the initial condition $x_0$. As an example:

$$x_3 = r^3 (1 - x_0)\, x_0\, (1 - r x_0 + r x_0^2)(1 - r^2 x_0 + r^2 x_0^2 + r^3 x_0^2 - 2 r^3 x_0^3 + r^3 x_0^4)$$

By varying the parameter *r* which represents the environmental conditions, a number of weird things happen (Figure 1), particularly for *r* equal or greater than 3.57, when periodic oscillations start being replaced by pure chaos.

In addition to sensitivity to the initial conditions, a second property of complex systems is indeed *deterministic chaos*: the underlying laws (physical, biological, etc.) may be orderly and even deterministic, yet chaotic behavior is possible.

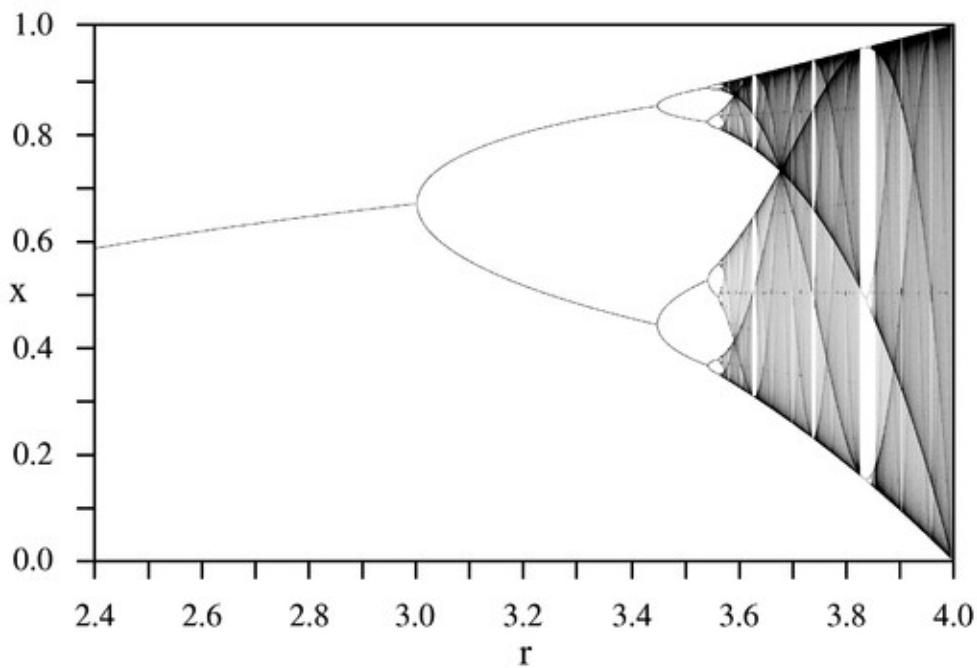

Figure 1: The logistic map

Iterating the logistic map is equivalent to applying geometric actions of stretching, bending and compression to the corresponding mathematical function. That is somewhat like kneading water and flour to make pizza: by repeated stretching and bending, two particles of flour dough that were very close at the beginning may end up being far apart; or a lump of dough that was initially concentrated may end up being evenly distributed. The Baker's Transformation (Ornstein, 1989), a formal expression of the pizza metaphor, is an abstract dynamical system that can be used as a paradigm for explaining the possibility of deterministic chaos. And since it can also serve as a mathematical model for coin tossing (*ibid*), it opens a way to delve into the profound nature of chaos and its possible relationship with randomness.

### 3.3 Non-linearity and emerging behavior

In addition to sensitivity to initial conditions and deterministic chaos, a third essential property of complexity is *emerging behavior*: even when the laws governing its components are well-known, a complex system may show a behavior that cannot be explained on those grounds.

A certain popular literature on complexity (Magrassi, 2013) tends to furnish us with examples from the living world or other high-level natural systems (flocks of birds and colonies of ants behave in ways inexplicable based on what we know of the capabilities of the individuals), however emergence has been known to physicists since Phil Anderson showed it in the case of groups of electrons in a semiconductor (Anderson, 1972). As he clarified many times afterward, emergent complex phenomena are not violations of the microscopic laws: they simply «do not appear as logically consequent» on them (Anderson, 1995).

Anderson's discovery of emergence in elementary particles closed the door to the hope of fully deciphering Nature by merely finding the fundamental laws of physics: it was the end of reductionism. Due to component interactions, at each geometrical level of Nature (quark, neutron, nucleus, atom, molecule, virus, life cell, etc.) new sets of laws may appear that, while compatible with the lower-level ones, introduce new knowledge.

## 4 The origin of complexity and its extent

What makes a system complex is not the number of components, but the interactions between them – the ultimate cause of non-linearity (Bridgman, 1927). This explains why complexity can emerge in "toy" problems / systems like Poincaré's three-body, May's prey-predator-food scenario or, to name one more recent finding, the coupling of a human heartbeat with breathing (Wessel et al., 2009).

A system made of many non-interacting parts is merely complicated (from Latin *complico*, to fold): it takes a long time to unfold it, to solve it, but it can be done step by step. On the other hand, a complex system, from *complector* (to encircle, to embrace firmly, to comprise, to unite under a single thought and a single denomination) (Magrassi, 2009b), is hard to tackle because, in addition to the laws governing the components, we need to study the system's overall behavior: the analytic approach must be complemented with the holistic one. Clearly, an increasing number of interacting components will give raise to increasing complexity, possibly exponentially: complex *and* complicated makes things worse.

This is the dynamical / systemic view of complexity. Another possible view is the computational / structural one: ultimately related to Gödel incompleteness theorems, the structural view is predominantly adopted in information theory, where it has to do with the computability of algorithms. The two views are ultimately connected via the concept of entropy (Ford, 1992; Falcioni et al., 2003), and, it should be noted, optimization problems can be described in terms of dynamical systems that become transiently chaotic as optimization hardness increases (Ercsey-Ravasz et al., 2011).

The discovery of emerging behavior in elementary particles (Anderson, 1972), along with the chaotic behavior seen in small deterministic systems, shows how intimately the natural world is permeated with complexity. For decades, however, this was never communicated effectively outside the physical community, despite Anderson gaining a Nobel Price in 1977 for related works. The news that usually make it out of the world of physics are those concerning the two extreme fields of elementary particles and astrophysics, because of the grandiose scale, the cost and the mediatic impact of projects such as super-accelerators or spacecraft-mounted probes and telescopes. News from other sub-domains of physical research rarely make it to life scientists or social scientists.

The *mesoscale* is that level of matter where (1) age is not much relevant, as it is instead for galaxies, and where (2) it is not useful to regard structures as groups of elementary particles (like in an atom), because these are far too many and statistical means or higher-level laws become necessary. This sub-domain of physics has always been a hotbed for powerful applications, like X-rays or transistors or lasers, but it was never regarded as a source of better explanations of Nature like it happened with subatomic physics or astrophysics. It is therefore not surprising that the consciousness of emergence as a physical phenomenon -as opposed to to an exclusive feature of living organisms- is taking a long time to make it to the mainstream of complex studies outside of physics or even outside of just condensed matter physics. Even the very notion that complexity reveal itself in minimal systems, often goes neglected.

## 5 Complexity and economics

The 21st century economic world is obviously characterized by an increasing number of connections. Financial markets are strongly interconnected. Economies are interconnected due to globalization (McKinsey, 2014). Enterprises are increasingly interconnected in supply and demand chains (Sodhi and Tang, 2010), ecosystems, and "clouds". Consumers are interconnected, and influence each others' behaviors, via communications forms of all sorts, such as social networks. This means that markets, prices, supplies, demands, consumers are all interacting and potentially giving raise to non-linear "systemic", emergent performance. All this could challenge the survival of economic agents or systems, making systemic risk highest (Malevergne and Sornette; 2006; Lo 2009; Battiston et al., 2011; Haldane and May, 2011; Haubrich and Lo, 2013; Haldane, 2014).

The suspect that non-linearity might be a factor at play in economics mounted among economists in the 1980s (Benhabib and Nishimura, 1979; Benhabib and Day, 1980; Day, 1983; Grandmont, 1983, 1985; Begg, 1984; Boldrin and Montrucchio, 1986; Van Der Ploeg, 1986), but it was eventually ruled out as a perturbation incapable of modifying the dominant economics paradigm. In the words of Jess Benhabib (Benhabib, 2008):

> *The aperiodic but bounded trajectories that characterize chaos and exhibit sensitive dependence on initial conditions cannot continue to diverge forever. They converge not to a point or a periodic cycle, but to a bounded chaotic or "strange" attractor. The dynamical system which induces the local separation and instability of the trajectories must eventually bend them back. The combination of local stretching and global folding generates the complex nature of the dynamics.*
>
> *Such dynamic behavior is in fact a familiar theme in economics that highlights the self-correcting nature of the economic system. Shortages create incentives for increased supply; dire necessities give rise to inventions as the invisible hand guides the allocation of resources. An equally familiar theme is that of instability: the multiplier interacts with the accelerator, leading to explosive or implosive investment expenditures; self-fulfilling expectations give rise to bubbles and crashes.*
>
> *In combination, these two themes suggest a nonlinear system, somewhat unstable at the core, but effectively contained further out. The contribution of the new literature on chaotic dynamics starting in the early 80s has been to demonstrate the compatibility of endogenous irregular fluctuations with equilibrium dynamics in economics.*
>
> *[...] At this point, while we know that standard dynamic equilibrium models with parameters calibrated to values often used in the literature may well generate chaotic dynamics, more definitive empirical evidence for chaos in economics has not yet been produced.*

Said economics paradigm is deeply rooted in linearity. A market is assumed to always be in the surroundings of an equilibrium (a point, line, surface, volume or hypervolume in state space), as well as to be efficient, unpredictable and capable of smoothing out all imperfections in the large numbers. Even before the "complexity awakening" of the 1980s, economists had long been aware of the limitations of such tranquil vision (Scarf, 1960; Sonnenschein, 1972; Stiglitz, 1975), but economic theory remains essentially a world of linear interactions and Gaussian-distributed risks: the same "disorganized complexity" (Weaver, 1948) of Brownian motion and equilibrium thermodynamics.

## 6 Econophysics

The prodromes of econophysics can be traced to L. Bachelier, V. Pareto, B. Manldelbrot or J. Tinbergen, and explicit econophysical suggestions were presented in the late 1970s (Rand, 1978; Ford,

1982). But it acquired momentum after the "Evolutionary Paths of the Global Economy Workshop", held September, 1987, in Santa Fe, New Mexico, where for the first time eminent economists, physicists and other natural scientists got together to discuss, among other things, the dominant paradigm in economics research and the possible contributions from physics and the natural sciences in general (Anderson et al. 1988).

These were the first attempts by physicists to step in and help, with the tools of their trade, proposing answers to questions such as «under what conditions chaotic behavior may emerge from a model incorporating specific economic structures» (Day, 1983), or «how large-scale patterns of catastrophic nature might evolve from a series of interactions on the smallest and increasingly larger scales, where the rules for the interactions are presumed identifiable and known» (ETH Zürich, 2019); and, perhaps, reducing «some perplexing results of economic theory [...] to a few elegant general principles with the help of some serious mathematics borrowed from the study of disordered materials» (Săvoiu and Simăn, 2013)

The "physics invasion" was encouraged by the growing availability of fine-grained empirical data in databases concerning markets and economic observations, on which scholars started applying mathematical methods typical of hard science and adopted an empirical approach consisting in looking at historical data without assuming much of any economic theory, while searching similarities with known phenomena in physics. It soon became possible for physicists to publish on economic modeling in physical journals, as was the case, e.g., with the pioneering works of Eugene Stanley and Rosario Mantegna in the early 1990s (Mantegna, 1991) (Mantegna and Stanley, 1995).

Non-linearity, systems operating far from equilibrium, and "organized disorder" (deterministic chaos, emerging behavior) were some of concepts brought in by the earliest econophysicists: just about the opposite of what happens in the rational-expectations- and efficient-market-hypothesis paradigm, where markets have no internal dynamics and chaos may only be stochastic. R. Mantegna, for example, was first to show "experimentally" that price indices in a Stock Exchange have statistical properties that are not Gaussian but rather compatible with a Lévy random walk with a non-local memory coupling price and time, and attributed the result to the openness and far-from-equilibrium characteristics of the system he had investigated, i.e. a financial market (Mantegna, 1991). Closed systems at equilibrium had been the economic modeling standard until then. In fact, Mantegna rightfully noted that physics, too, was abusing such reassuring assumptions: «an improper generalization of a property of the best known stochastic process, [...] Gaussian Brownian motion […]», had caused «a lack of interest for Lévy stable distributions in physics», despite the fact that «widespread investigations of open non-equilibrium systems» had shown that «statistical distributions with power-law tails are often present in natural and social phenomena» (*ibid*).

The dissipative systems that Mantegna had in mind were similar to those that had been studied in the thermodynamic approach to complexity, inaugurated by I. Prigogine (Prigogine, 1977) and that subsequently generated a strand of econophysical study sometimes called thermoeconomics. Another early econophysical approach consisted in drawing analogies from models developed in condensed matter physics. For example, a spin glass is a system characterized by non-ergodicity, an extreme fragility with respect to small changes in parameters and the de facto absence of equilibrium (Anderson 1972b; Parisi, 1983): this was one of the first models proposed in econophysics (Anderson et al., 1988; Bouchaud, 2009).

The widest class of models are agent-based models, which econophysics shares with, and in fact imported from, pure economics research. Here, the domain under investigation is explored via computer-based simulation. Numerical investigation of a model does not rigorously prove anything, yet provides a formidable tool, a «telescope of the mind multiplying human powers of analysis and insight just as a telescope does our powers of vision» (Buchanan, 2008). Simulations of this sort often lead to situations very different from the perpetual quasi-equilibrium of efficient markets. For example, catastrophic meltdowns can take place abruptly, something that in incumbent macroeconomic models, and contrary to empirical evidence, may happen only with infinitely small probabilities. The earliest of this kind of simulations by econophysicists are to be found in, e.g., (Lux and Marchesi, 1999; Macal et al., 2004), and had been anticipated by purely economics research such as in (Kim and Markowitz, 1989) or (Arthur et al., 1997), a work which counted as co-author John Holland, the "father" of genetic algorithms (Holland, 1975).

Deep and up-to-date accounts of this thread of research as well as the whole of econophysics will be found in (Aste and Di Matteo, 2010; Kutner et al., 2018), while a more synthetic overview of the major achievements is in (Bouchaud, 2019). On the other hand, the struggle by econophysicists to get recognition within the economics community is narrated in (Gallegati et al, 2006; Ball, 2006; Omerod, 2016; Ausloos et al., 2018).

In a nutshell: (1) the cooperation between econophysicists and economists is still very low; (2) econophysics results have not shocked the economics community; (3) econophysicists resist to accepting an overarching macroeconomic model of «mild fluctuations around a stable equilibrium» (Bouchaud, 2019); (4) nobody knows how to reconcile the supposed (by mainstream economists) Brownian motion and ergodicity of financial markets with the intermittent, fat-tailed nature of asset prices, although the issue started being investigated long before econophysics (Mitchell, 1915; Olivier, 1926; Mills, 1927; Larson, 1960; Houthakker, 1961); (5) econophysicists have a tendency to erroneously assimilate the efficient-markets hypothesis (EMH) with the presumed Gaussian character of stock market variations (Ausloos et al., 2018).

The latter issue is particularly tricky. The EMH has gained among econophysicists the fame of a dogma -a word that actually recurs often in their literature. However the true meaning of the theory is sometimes dismissed too soon in such papers. In Fama's canonical definition, a market is said to be efficient with respect to an information set if the asset price "fully reflects" that information set (Fama, 1970). I.e.,

$$E ( p_{j,t+1} | \Phi_t ) = p_{j,t} [ 1 + E ( r_{j,t+1} | \Phi_t ) ]$$

where E is the expected-value operator; the random variable $p$ is the price of asset j at time t; $r$, also a random variable, is the one-period percentage return; $\Phi_t$ is a symbol for whatever set of information is assumed to be "fully reflected" in the price at t. As (Sewell, 2011) effectively put it, the definitional "fully" is an exacting requirement, suggesting that no real-world market could ever be efficient. This implies that the EMH may be exactly false although asymptotically true. This is not an uncommon situation in science. To use an analogy drawn from a field adjacent to the present discussion, a nonlinear dynamical system has zero probability to sit exactly on a chaotic attractor in phase space, however all its trajectories are asymptotic to some generic trajectory on the attractor with probability 1, and thus the dynamics of the attractor still governs the observed long-term behavior of the system (Ornstein, 1989).

The reason the EMH is sometimes labeled a dogma is that it still cannot be tested. However no alternative theory so far has either been tested or even formally proved. When the criticism was raised that an efficient-markets-based theory would always be clueless at anticipating shocks such as the subprime crisis culminated in the 2008 financial meltdown, Eugene Fama replied that the 2008 financial crash was the efficient-market anticipation of an upcoming economic recession, and that finance had been the victim, not the cause, of the economy meltdown (Cassidy, 2010).

### 7 The effects of complexity

The merely approximate correctness of the EMH also leads to an epistemological issue which is related to the role of complexity in economics. Many things in science are approximate, and no physical laws are universally exhaustive.

The basic physical constants, such as Plank's, Boltzmann's and Avogadro's, or the speed of light and the electron charge, that is numbers playing key roles in fundamental equations, have always been known with approximation (they were given de jure exact values effective May 20, 2019, in order to rationalize measurement in science, technology and everyday life), but this has not prevented science and technology from making great progress, exactly like it happened with the linearity assumption.

Objects in the mesoscale do not seem to obey [only] the laws of quantum mechanics, and classical physics "emerges" -most likely due to quantum decoherence- out of a quantistic underlying world. Heisenberg's uncertainty principle is negligible and uninteresting when the energy of the objects being observed is much larger than Plank's constant, i.e. about $10^{-34}$ Joule*second. In many everyday situations, including some highly sophisticated technologies, we are not concerned about the effects of Special or General Relativity, because the objects we deal with do not usually move at speeds close to light's or travel extremely long distances: the relativistic effects are negligible most of the time.

Our "laws of nature" are imperfect and only work within roughly predefined bounds. Thinking of the economy, one may therefore ask: (1) Is the EMH's asymptotic [presumed] validity really a problem? Does it always have a measurable, non-negligible impact on applications, such as the S&P500 or Milan's Stock Exchange (the object of Mantegna's famous analysis)? Are irregularities unforeseen by the EMH really due to the EHM's essential invalidity, or do they originate from emerging higher-level laws which we still ignore? And (2) does the non-linearity of economics always have a measurable, non-negligible impact?

Neither the economics or the econophysics scientific literature seem to contain answers to those questions.

### 8 Conclusions

The natural sciences are still struggling with the complexity challenge that stems from relaxing linear assumptions when modeling dynamical systems, like Fermi, Pasta, Ulam and Tsingou-Menzel first did in 1953 (Dauxois and Ruffo, 2008). Contrary to a widespread belief, the process is far from concluded. Physics itself is still largely based on linear models at its core, and some of the key results of non-linear studies, obtained in chemistry and condensed-matter physics, still have not made it to the whole of the natural sciences themselves, much less to economics.

Considering the quasi-obvious "complexity" (in the sense precisely defined in this paper) of the macroeconomy and of global financial markets, as well as the scarcity of economists who have participated to the econophysics endeavor so far, perhaps economics research would benefit if physicists and economists finally joined to make econophysics a hotbed for non-linear studies in economics.

This would possibly entail a more in-depth exploration of complexity than it has been conducted thus far in econophysics. It may be necessary to concentrate less on the applications than on the basics of economic complexity, beginning with expansion and deepening of the study of small systems with few interacting components, while until thus far complexity has been assumed to be a prerogative of complicated systems only. It is possible that without a thorough analysis at that level, the understanding of systems that are at the same time complex and complicated will continue to elude us.

It might as well be useful to deepen from an economics viewpoint the study of the entanglement between stochastic and chaotic processes, like it happens in statistical physics, chemistry and information theory: which means that, perhaps, some scholars from all those disciplines should join the econophysics community, whatever its future name is going to be.